\documentclass{article}
\usepackage[english]{babel}

\usepackage{lettrine}
\usepackage{amsmath}

\usepackage[colorlinks=true, allcolors=blue]{hyperref}
\usepackage[a4paper,left=20mm,right=20mm,top=15mm,bottom=20mm]{geometry} 
\usepackage{cite}
\usepackage{graphicx}
\usepackage{epstopdf}
\usepackage{fancyhdr}
\usepackage[T1]{fontenc}
\usepackage[utf8]{inputenc}
\usepackage{authblk}
\usepackage{multirow}
\usepackage{booktabs}

\newcommand{\figcaption}{\def\@captype{Fig. }} 
    {} 
\makeatother

\title{Self-Amplification-Assisted Highly Efficient Integrated Laser}

\author[1,$^\dagger$]{Jiangwei Wu}
\author[1,$^\dagger$]{Xiongshuo Yan}
\author[1]{Xueyi Wang}
\author[1]{Tingge Yuan}
\author[1]{Chengyu Chen}
\author[1,2,*]{\\Yuping Chen}
\author[1,3,4]{Xianfeng Chen}

\affil[1]{School of Physics and Astronomy,State Key Laboratory of Advanced Optical Communication Systems and Networks, Shanghai Jiao Tong University, 800 Dongchuan Road, Shanghai 200240, China}
\affil[2]{School of Physics, Ningxia University, Yinchuan, 750021, China}
\affil[3]{Shanghai Research Center for Quantum Sciences, Shanghai 201315, China}
\affil[4]{Collaborative Innovation Center of Light Manipulations and Applications, Shandong Normal University, Jinan 250358, China}

\affil[ ]{ }

\affil[ ]{Jiangwei Wu: wjw2016@sjtu.edu.cn}
\affil[ ]{Xiongshuo Yan:  xiongshuoyan@sjtu.edu.cn}
\affil[ ]{Xueyi Wang: wangxytrish@sjtu.edu.cn}
\affil[ ]{Tingge Yuan: yuantg0516@sjtu.edu.cn}
\affil[ ]{Chengyu Chen: chengyuchen@sjtu.edu.cn}
\affil[ ]{Hao Li: lihaosky@sjtu.edu.cn}
\affil[ ]{Xianfeng Chen: xfchen@sjtu.edu.cn}
\affil[ ]{ }

\affil[ ]{Yuping Chen$^*$: ypchen@sjtu.edu.cn}
\affil[ ]{School of Physics and Astronomy, Shanghai Jiao Tong University,}
\affil[ ]{800 Dongchuan Road, Shanghai 200240, China}
\affil[ ]{Tel: +86-13816373910}
\affil[ ]{ }

\affil[$^\dagger$]{These authors contributed equally to this work.}

\date{} 

\begin{document}
\maketitle
\newpage
\begin{abstract}

Light source is indispensable component in on-chip system. Compared with hybrid or heterogeneous integrated laser, monolithically integrated laser is  more suitable for high density photonic integrated circuit (PIC) since the capability of large-scale manufacturing, lower active-passive coupling loss and less test complexity. Recent years have seen the spark of researches on rare-earth ion doped thin film lithium niobate (REI:TFLN), demonstrations have been made both in classical and quantum chips. However, low output power and limited quantum emitting efficiency hinder the application of the chip-scale laser source based on REI:TFLN. Here a highly efficient integrated laser assisted by cascaded amplifiers is proposed and experimentally prepared on Erbium-doped TFLN. A slope efficiency of 0.43$\%$ and a linewidth of 47.86 kHz are obtained. The maximum integrated laser power is 7.989 µW. Our results show a viable solution to improve efficiency by self-amplification without changing the intrinsic quantum emitting efficiency of the material, and our design has potential application in  incorporating with functional devices such as optical communications, integrated quantum memory and quantum emission.

\end{abstract}

\section*{Introduction}
Integration is of great significance in device miniaturization and energy efficiency improvement \cite{bradley2011erbium}. PIC as one of the important goals for the development of photonics have attracted enormous attentions and became one of the most investigated research fields. The light source is the heart of the optical system and the beginning of any optical application \cite{jiang2016whispering, zhang2017advances,zhang2014room, cao2020reconfigurable,tang2021laser, zhang2021halide}. Many solutions of lasers have been provided. The requirements of an ideal chip-scale lasers include sufficient large power with high power efficiency, continuous-wave emission in fiber-based communication band, compatibility with CMOS processes and so on \cite{zhou2023prospects}.  Taking silicon-based PIC as an example. III-V lasers have been commercialized for a long time and  have been explored extensively to combine with silicon photonics using hybrid and heterogeneous integration \cite{zhang2020low, shang2022electrically, wei2023monolithic,alkhazraji2023linewidth,remis2023unlocking,ma2023room}. Meanwhile, silicon-based Raman lasers \cite{ferrara2020integrated}, silicon quantum dots lasers \cite{dohnalova2014silicon} and other Group-IV lasers that could be monolithically integrated with silicon photonics are attracting researchers’ attention.

Lithium niobate (LN) is another emerging integrated photonic platform material. It has been widely used in optical and microwave fields, due to its rich properties such as wide transparent wavelength range, excellent electro-optic, acousto-optic characteristics and large second order nonlinear susceptibility \cite{nikogosyan2006nonlinear}. In particular, since the TFLN has been commercially available, many compact and low-cost photonic devices with high performance can be achieved on this platform \cite{2017Monolithic,boes2018status,lin2020advances}, such as efficient frequency convertors \cite{luo2018highly, ye2020sum, ge2018broadband, lin2016phase, lin2019broadband,li2022efficient}, electro-optical modulators \cite{wang2018integrated, li2020lithium, xu2020high, pan2020first,zhu2022spectral}, acousto-optic modulator \cite{wan2022highly} and frequency comb sources \cite{wang2019monolithic, fang2019efficient, 2019Self}. As for integrated laser on LN, solutions are limited. Considering the stringent requirements, on-chip lasers based on LN await further investigation. The hybrid integrated electrical pumped laser with TFLN has been demonstrated \cite{shams2022electrically}. However, monolithically integrated laser is preferred since the capability of large-scale manufacturing, lower active-passive coupling loss and less test complexity. Doping rare earth ions into  LN crystal could make it a gain medium, which has been performed on many materials such as silica fiber, silicon nitride and yttrium aluminum garnet \cite{agazzi2013energy, mu2020high, ronn2020erbium, vazquez2014erbium, ronn2019ultra, min2004erbium,li2023optically}. With a doped TFLN, a monolithically integrated laser is now possible. We developed an Erbium (Er) doped TFLN and fabricated a microdisk laser for the first time in the world in 2021 \cite{liu2021chip}.  REI:TFLN has been considered as a promising material platform for both classical and quantum integrated photonics \cite{jia2022integrated, chen2022photonic}. In recent years, many active devices for classical application including microdisk laser, microring laser, single-mode laser and waveguide amplifier were reported on this platform \cite{liu2021chip,yin2021electro,wang2021chip,luo2021microdisk,luo2021chip,liu2021tunable,gao2021chip,zhang2021integrated,xiao2021single,li2021single,lin2022electro,liang2022monolithic,zhu2022electro,guan2024monolithically,yu2023chip,zhou2023monolithically, yan2021integrated,zhou2021chip,chen2021efficient,luo2021onchip,liang2022high,zhang2022chip,cai2021erbium}. Quantum applications like single photon sources and quantum memory have also been demonstrated \cite{saravi2021lithium, yang2023controlling, xia2022tunable}. 

The laser transition of trivalent Er ions matches the C-band telecom window, making it the most attractive dopant ion for photonic application. As for Er-doped on-chip laser, the output laser powers of most existed results are too small to support subsequent work, and their slope efficiencies are low ($\sim10^{-4}$), so that the pump powers have to be large. We propose and experimentally demonstrate an efficient integrated amplifier-assisted laser (IAL) on Er-doped TFLN, which consists of a microdisk resonator and a spiral waveguide in a relatively small footprint ($\sim$0.405 mm$^2$). A maximum on-chip laser power of 7.989 µW is achieved under non-tunable pump, while with a tunable pump source, the IAL could emit 7.26 µW-laser with a slope efficiency is $4.3\times10^{-3}$, and laser linewidth is 47.86 kHz. We also examine the thermal tunability of the device. The numerical simulation is given and it fits well with the experimental data. The IAL would be a possible solution for efficient monolithically integrated light source in large scale PIC and enable more types of quantum applications \cite{saglamyurek2011broadband, sinclair2016proposal,sinclair2017properties, doi:10.1126/sciadv.adf4587}.
  
\section*{Results}

\begin{figure}[h]
\centerline{\includegraphics[width=0.8\textwidth]{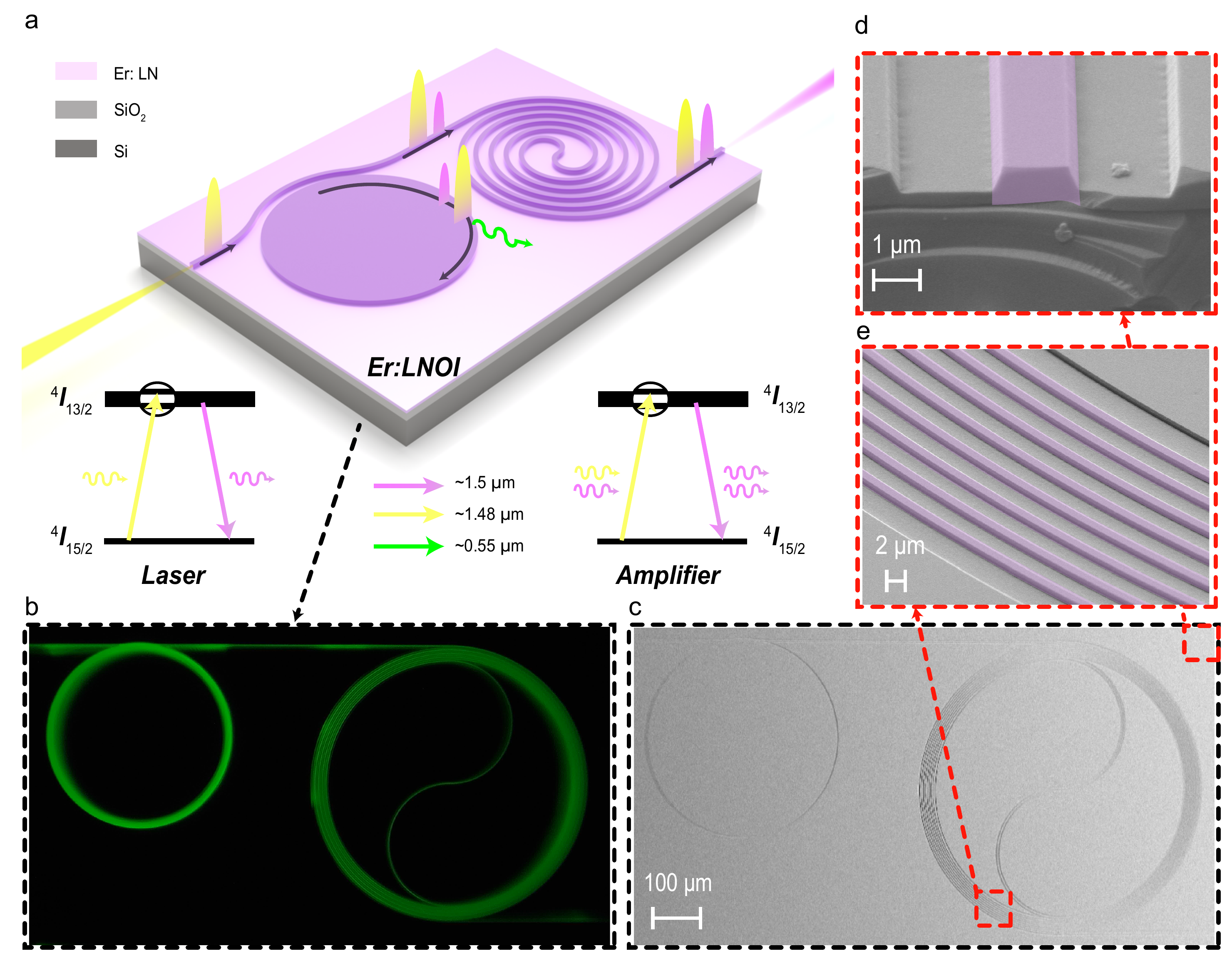}} 
\caption{\textbf{Concept and structure of the IAL.} \textbf{a} Schematic of the IAL. \textbf{b} Microscopic images of the IAL with pump.  \textbf{c, d} and \textbf{e} Scanning electron microscope images of  the IAL, zoomed-in figures are the edge coupler and waveguide of amplifier. } 
\label{Fig. 1} 
\end{figure}

The structure of the IAL is shown in Fig. \ref{Fig. 1}a. When the pump laser is injected in the waveguide through an edge coupler and coupled into the microdisk resonator, the Er ions are excited and then emit C-band laser. As the signal laser and pump laser couple back and propagate along the spiral waveguide amplifier, energy is further transferred into the signal laser. The energy level diagrams included in Fig. \ref{Fig. 1}a plot the lasing and amplification processes respectively. It is worth noting that the cooperation up-conversion of Er ions under pump would also generate a green light. This process is detrimental for quantum emitting efficiency of C-band laser, while it could help us to estimate the power density of the pump laser. Fig. \ref{Fig. 1}b shows the microscopic images of the device with pump on. The green emission highlights the microdisk cavity and the spiral waveguide, where the enhancement of light-matter interaction in the resonator is obvious.

\begin{figure}[h]
\centerline{\includegraphics[width=0.8\textwidth]{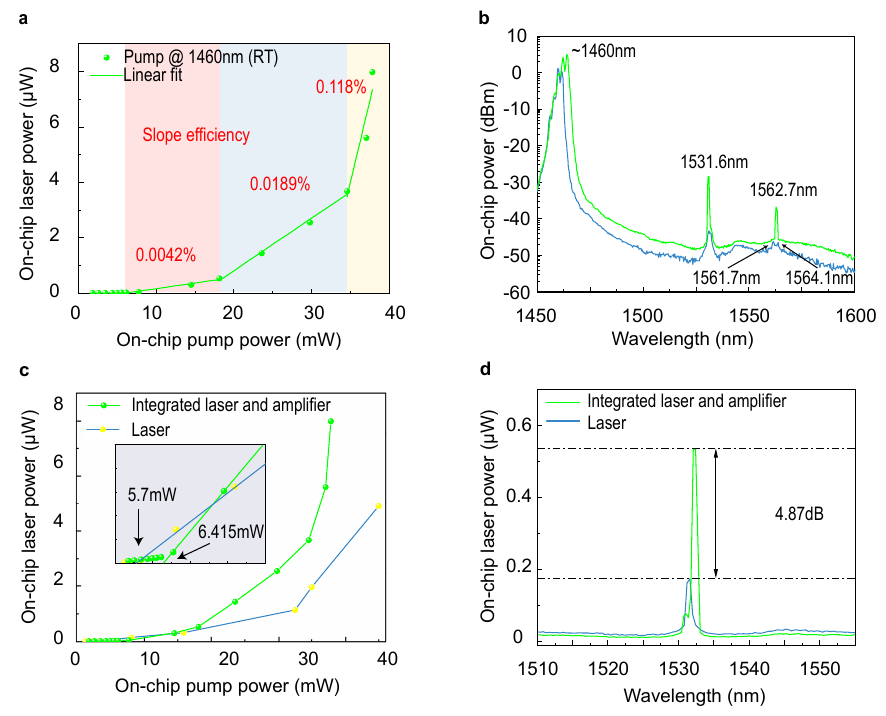}} 
\caption{\textbf{Performance under non-tunable pump} \textbf{a} Relationship between laser power and non-tunable pump power. \textbf{b} The laser spectra of the IAL under non-tunable pump, blue: 7.37 mW, Green: 23.58 mW. \textbf{c} Relationship between laser power and non-tunable pump power of the microdisk laser and IAL, inset shows the relationships around laser threshold.  \textbf{d} Spectrum around 1531.6-nm laser from laser and the IAL with same non-tunable pump at 17.98 mW.} 
\label{Fig. 2} 
\end{figure}

The scanning electron microscope image of the device and zoomed-in figures of the edge coupler and spiral waveguides are shown in Fig. \ref{Fig. 1}c, d and e. We chose a 300-µm-diameter microdisk as the laser resonator, so that the FSR of the cavity is small enough for a non-tunable pump laser to couple into. The width of the waveguide is set as 1 µm for single mode propagation, and the microdisk-waveguide coupling region is adiabatically tapered to 0.6 µm for evanescent coupling. As for the spiral waveguide, the minimum bending radius is 100 µm to reduce scattering loss, and the gap between two adjacent waveguides is 2 µm to avoid crosstalk. The whole device has a footprint of $\sim$0.405 mm$^2$.
The IAL is first pumped by a non-tunable 1460-nm laser, whose linewidth is several nanometers. The properties are illustrated in Fig. \ref{Fig. 2}. The relation between laser and pump power is shown in Fig. \ref{Fig. 2}a. To compare the device with single microdisk laser and single spiral amplifier, we use on-chip power in this essay to eliminate the effect of coupling difference. The laser power increases with enlarged pump power in a changing slope efficiency, as shown in differently colored blocks in Fig. \ref{Fig. 2}a. The nonuniform efficiency might be originated from the nature of IAL. When coupled pump power exceeds the laser threshold, the device behaves like a single microdisk laser at first, since the ground state absorption (GSA) dominates in the spiral waveguide region at that time. As pump further increases, more Er ions could be excited so the amplifier becomes gainful. The maximum slope efficiency of 0.118$\%$ is achieved when gain of the amplifier saturated, meanwhile a maximum on-chip laser power as high as 7.989 µW is observed. (The detailed characterizations of single microdisk laser and single spiral amplifier are shown in the Supplementary Materials.) Apart from signal around 1531 nm, laser around 1562 nm could also be generated since the microdisk cavity supports multi-mode laser emission. Fig. \ref{Fig. 2}b shows the collected spectra covering both pump and main laser wavelengths with different pump power, the multiple peaks around 1562 nm also denote multi-mode emission.
The lasing performance of the IAL is compared with the single microdisk laser in Fig. \ref{Fig. 2}c. The two devices are fabricated on the same chip simultaneously, and the microdisk cavities share same geometrical parameters including microdisk resonator radius and the coupling gap with bus waveguide. The coupling loss caused by edge coupler is excluded to get an on-chip pump power. The IAL possesses a larger threshold power of 6.415 mW (the inset of Fig. \ref{Fig. 2}c). Because longer waveguide with larger GSA needs more pump power to be transparent for laser signal to output. So that the slope efficiency of the microdisk laser is slightly larger than that of IAL with small pump power, and soon be surpassed with increased pump power. While the two devices are under same on-chip pump power of 18 mW, we achieve a laser intensity enhancement of 4.87 dB as shown in Fig. \ref{Fig. 2}d. The mismatch of the two signal peaks might be due to the size deviation of the cavity caused by fabrication error, so they support shifted resonances. These results suggest that the IAL exhibits superior laser output performance to single microdisk laser and thus could be a better candidate for on-chip integration.

\begin{figure}[h]
\centerline{\includegraphics[width=0.8\textwidth]{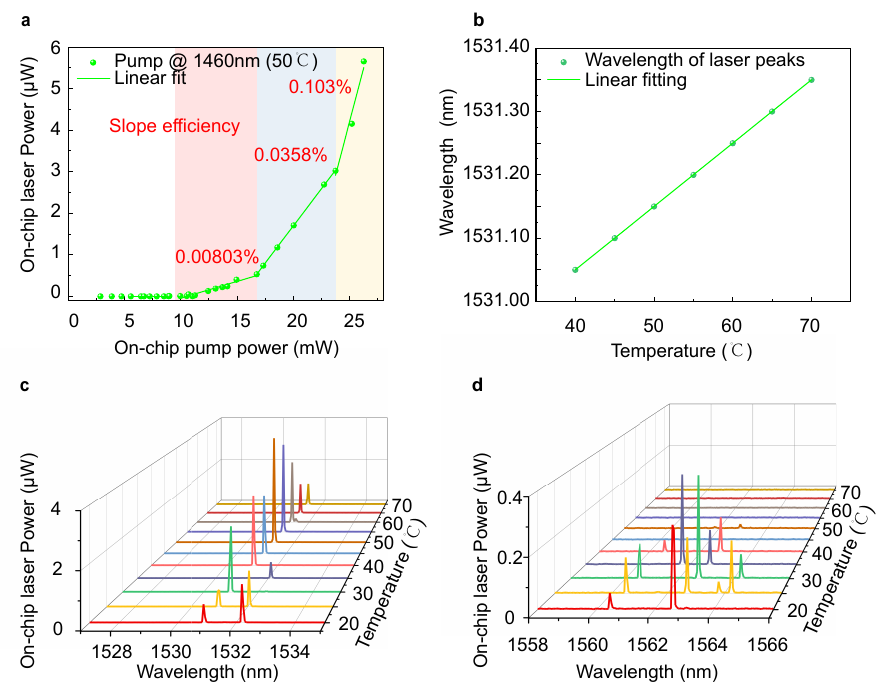}} 
\caption{\textbf{Temperature properties of the IAL.} \textbf{a} 1531.2-nm laser vs. 1460-nm pump relationship of the IAL at 50 $^\text{o}$C. \textbf{b} The relationship between the wavelength of laser peaks and temperature (from 40 $^\text{o}$C to 70 $^\text{o}$C). \textbf{c} Laser mode (around 1530 nm) evolution with temperature increasing from 20 $^\text{o}$C to 70 $^\text{o}$C. \textbf{d} Laser mode (around 1562 nm) evolution with temperature increasing from 20 $^\text{o}$C to 70 $^\text{o}$C.} 
\label{Fig. 3} 
\end{figure}

We also examine the performance of the IAL at different temperatures to prove the stability under various conditions and the thermal tunability. Fig. \ref{Fig. 3}a shows that the IAL behaves similarly at 50 $^\text{o}$C as that at room temperature. There is no significant difference in its slope efficiency and the maximum on-chip laser power at different temperature. The signal peaks are shifting with the temperature changing from 40 $^\text{o}$C to 70 $^\text{o}$C, as shown in Fig. \ref{Fig. 3}b, indicating a thermal tuning coefficient of 10 pm/$^\text{o}$C. It is worth noting that the mode evolution at the two main lasing bands (around 1531 nm and around 1562 nm) with temperature increasing could be observed in Fig. \ref{Fig. 3}c and Fig. \ref{Fig. 3}d. Apart from the thermal tuning effect, multimode lasing becomes single mode lasing (around 1531 nm) when the chip is heated. And the multi-mode lasing at around 1562 nm is suppressed at high temperature (>50 $^\text{o}$C). These phenomena are caused by the overlap shifting of cavity resonance and the gain profile of the Er ions. And it provides a potential degree of freedom to manipulate the device.

\begin{figure}[h]
\centerline{\includegraphics[width=0.8\textwidth]{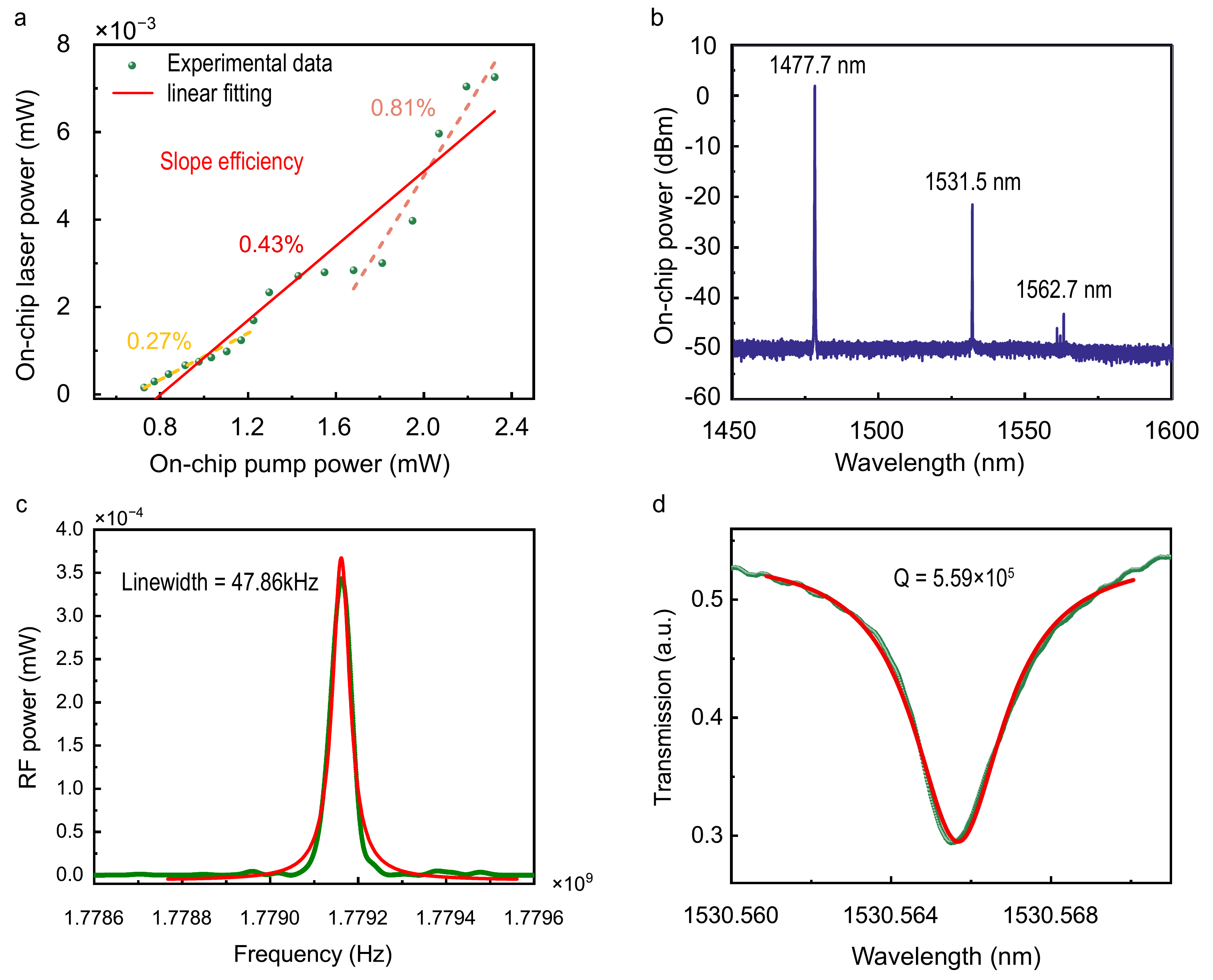}} 
\caption{\textbf{Performance under tunable pump.} \textbf{a} Relationship between laser power and tunable pump power. \textbf{b} The collected spectrum of the IAL under tunable pump. \textbf{c} The linewidth of the signal laser. \textbf{d} Lorentzian fitting of a measured mode around 1530.56 nm exhibiting a loaded Q factor of $5.59\times10^5$.} 
\label{Fig. 4} 
\end{figure}

Compared with non-tunable laser, tunable laser is more expensive and often with larger volume. But the narrow linewidth laser emission and tunability would help examine the optimum performance of the IAL. We use a tunable 1480-nm laser as the pump and obtain the light-in-light-out (L-L) relation as shown in Fig. \ref{Fig. 4}a. The wavelength of pump laser is tuned to achieve best coupling and oscillating condition. The behavior of changing slope efficiency is similar with that under non-tunable pump. The average slope efficiency is 0.43$\%$, while in some region, as fitted with yellow and pink dashed lines, the slope efficiencies are 0.27$\%$ and 0.81$\%$, respectively. The maximum laser power is 7.26 µW, slightly lower than that with non-tunable pump since the pump power is limited. The higher efficiency is easily to explain. Fig. \ref{Fig. 4}b shows the collected spectrum covering both pump and main laser wavelengths. Compared with non-tunable pump laser (shown in Fig. \ref{Fig. 2}b), the linewidth of tunable pump laser is quite narrow, thus the power needed to excite the Er ions is reduced. The lowered pump power makes it easier to perform a linewidth measurement. The output of the IAL is combined with another C-band tunable laser, the beat note signal is collected and analyzed using an electrical spectrum analyzer (ESA). The linewidth of the output laser at 1531.5nm is 47.86 kHz, as shown in Fig. \ref{Fig. 4}c.
The propagation loss is derived by evaluating the intrinsic Q of a simultaneously fabricated microring with 200-nm radius which is similar to that of the spiral waveguide. The resonance at around 1531 nm is shown in Fig. \ref{Fig. 4}d with a loaded Q factor of 5.59 × 10$^5$, and the calculated intrinsic Q is 6.24 × 10$^5$. The propagation loss $\alpha$ is calculated by:
\begin{equation}
    \alpha=\frac{2\pi N_{eff}}{\lambda_0Q_i},
\end{equation}
where $N_{eff}$ is the effective refractive index of the waveguide at the wavelength $\lambda_0$. The obtained propagation loss around 1531 nm is 0.53 dB/cm and it is then used to estimate the gain performance of the IAL.

\section*{Discussion}
\begin{figure}[h]
\centerline{\includegraphics[width=0.8\textwidth]{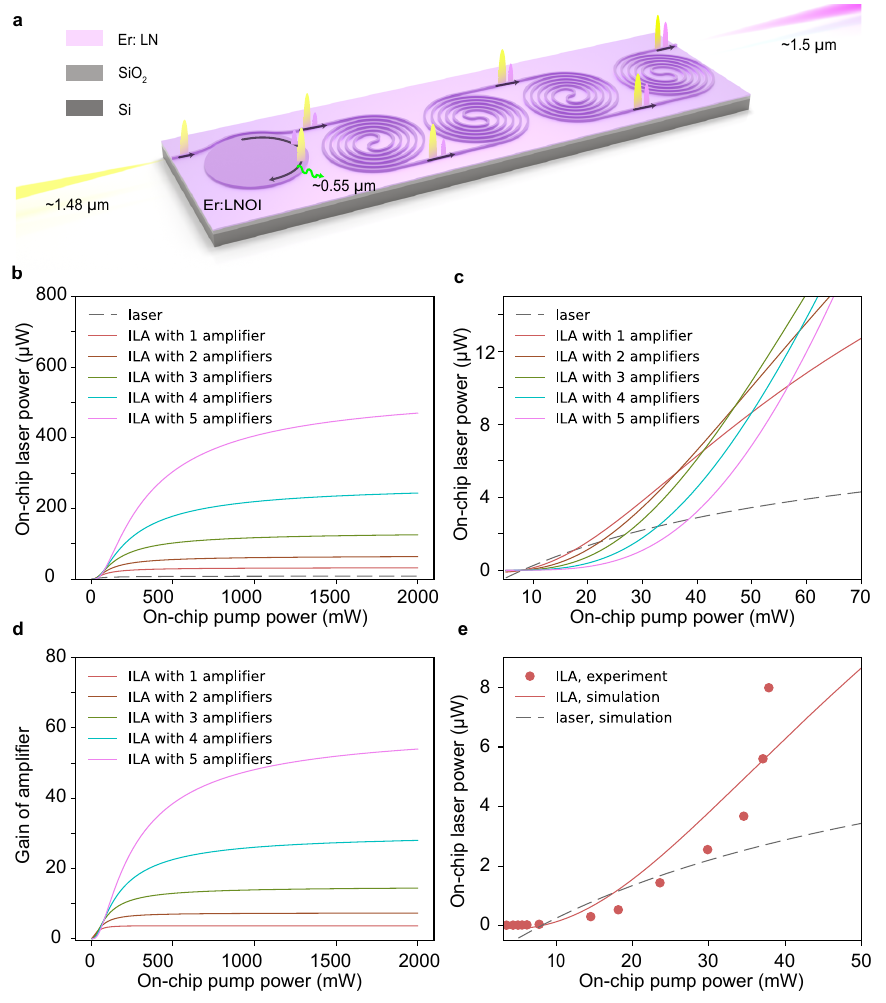}} 
\caption{\textbf{Numerical simulation.} \textbf{a} Schematic of the cascaded IAL. \textbf{b} Simulated laser power from the cascaded IAL with different number of spiral waveguide amplifiers. \textbf{c} Zoomed-in figure of \textbf{b}. \textbf{d} The gain of the amplifier in the cascaded IAL. \textbf{e} Comparison of the simulation and experiment result of the IAL.} 
\label{Fig. 5} 
\end{figure}

The gain saturation of amplifier limits the maximum on-chip laser power. For a waveguide amplifier, the gain saturation is mainly caused by pump absorption saturation of the Er ions. Increasing the doping concentration of Er ions might achieve higher gain, while the energy-transfer upconversion among neighboring Er ions and fast quenching would be severe with high Er concentration, which would lower the pump efficiency. Prolonging the waveguide might be a safer solution for high power laser output.
Fig. \ref{Fig. 5}a illustrates the IAL with multiple spiral waveguides, i.e., cascaded IAL which could increase the length of amplifier while add a small footprint. We perform a numerical simulation to predict the maximum laser output of a cascaded IAL, and the results are shown in Fig. \ref{Fig. 5}b and Fig. \ref{Fig. 5}c. The dashed line denotes the simulated microcavity laser output, which is set as the signal laser in the calculation of the cascaded IAL output. The approximation of uniform excitation along the waveguide is taken, and gain coefficient behavior with large signal and under large pump power are considered in the calculation. As the number of spiral amplifiers increased, the threshold would increase and the maximum output laser power is increased. Maximum laser output in the order of one hundred microwatts could be achieved in the IAL with 5 cascaded spiral waveguides, while larger pump power is needed to compensate the increased loss, as shown in Fig. \ref{Fig. 5}c. The gain of the amplifier, $G$ is defined as:
\begin{equation}
    G = \exp[(g+\alpha)l],
\end{equation}
where 
\begin{equation}
    g\propto\frac{g_0}{[1+\frac{I_{psa(l)}}{I_p}](1+\frac{I_l}{I_{ls}})},
\end{equation}
$\alpha$ is the loss of the amplifier, and $l$ is the length of the amplifier. $g_0$ is the gain coefficient of the amplifier with small signal,  $I_p$ is the pump power, and $I_{psa(l)}$ is the saturated pump power of the amplifier, which depends on the length of waveguide. $I_l$ and $I_{ls}$ are the laser output of the microcavity and the saturated laser power respectively. Fig. \ref{Fig. 5}d shows the relationship between pump power and the gain of the amplifier, where the saturation occurs in high pump. The experimental result fits well with the simulation as shown in Fig. \ref{Fig. 5}e, minor deviation might come from the power-dependant loss coefficient. 

Table 1 \ref{Table. 1} compares the typical parameters of reported works on Er-doped on-chip laser. Our devices exhibit the most balanced performance considering the compactness, output power, slope efficiency and linewidth. For  further improvement in those figure of merits, multiple methods could be applied. The efficiency could be higher  by co-doping with Ytterbium ions. And bi-directional pumping could achieve larger output laser power.  Optimization of the dry etching would result in higher loaded quality factor and narrower  linewidth. 
The polarization of applied pump laser has an apparent influence to the signal laser both in power and wavelength. The mode-field-matching condition between the lens fiber and the edge coupler accounts for the coupling loss of pump laser, and consequently the on-chip pump laser power. For a single mode ridge waveguide, the mode distribution of TE and TM modes are different, thus the coupling efficiency with lens fiber varies. On the other hand, since LN is an anisotropic material, the absorption and emission cross sections of the doped Er ions are polarization-dependent \cite{huang1994evaluation}. In our experimental setup, a polarization controller is used to change the applied polarization of pump laser, and the mode competition of output signal laser at telecommunication band are observed with different polarization.

\begin{table}[h]
\begin{center}
\resizebox{1\columnwidth}{!}{
    \begin{tabular}{ccccccc}
        \toprule
        \multirow{2}*{Device} & \multirow{2}*{Footprint} &\multirow{2}*{Pump $\lambda$ } &\multicolumn{3}{c}{Signal}&\multirow{2}*{Ref.}\\
        \cmidrule{4-6}
        &  & ($nm$)& Power ($nW$) &Slope efficiency & Linewidth & \\ 
        \midrule
        Disk & d = 150 µm & 1460 & $\sim 500$ & $3.15\times10^{-5}$ & 0.12 nm & \cite{liu2021chip}\\
        Racetrack ring & 400$\times$ 900 µm$^2$ & 980 & $\sim 35$ & $4.38\times10^{-5}$ & N/A & \cite{yin2021electro}\\
        Disk & d = 200 µm & 976 & $\sim 140$ & $1.92\times10^{-4}$ & 24 pm &\cite{wang2021chip}\\
        Disk & d = 90 µm & 974 & $\sim 0.4$ & $6.5\times10^{-7}$ & N/A &  \cite{luo2021microdisk}\\
        Ring & d = 100 µm & 974 & $\sim 0.1$ & $6.61\times10^{-7}$ & 10 pm & \cite{luo2021chip}\\
        Coupled disk and ring & d$_{1,2}$= 150, 165 µm & 974 & $\sim 95$ & $4.41\times10^{-5}$ & N/A & \cite{liu2021tunable}\\
        Coupled disk & d$_{1,2}$ = 29.8, 23.1 µm & 977.7 & $\sim 50$ & $7.0\times10^{-5}$ & 348 $kHz$ &\cite{gao2021chip}\\
        Coupled ring & d$_{1,2}$= 170, 200 µm & 979.6 & $\sim 40$ & N/A & 5 pm & \cite{zhang2021integrated}\\
        Coupled ring & d = 200 µm & 1484 & 310 & $1.45\times10^{-4}$ & 1.2 MHz & \cite{xiao2021single}\\
        Ring & d = 210 µm & 1484 & $2.1\times10^3$ & $1.2\times10^{-4}$ & 0.9 MHz & \cite{li2021single}\\
        Disk & d = 29.8 µm & 968 & $2\times10^3$ & $1.0\times10^{-4}$ & 322 Hz & \cite{lin2022electro}\\
        Ring & d = 400 µm & 976 & $\sim 100$ & $8.33\times10^{-6}$ & 45 pm & \cite{liang2022monolithic}\\
        Disk & d = 200 µm & 976 & $\sim 10$ & $\sim5\times10^{-6}$ & N/A & \cite{zhu2022electro}\\
        Disk & d = 1 mm & 980 & $62.1\times10^3$ &  $5.96\times10^{-3}$ & 0.11 MHz & \cite{guan2024monolithically}\\
        Disk with spiral waveguide & 450$\times$900 µm$^2$ & 1480 & $7.26\times10^3$ & $4.3\times10^{-3}$ & 47.86 kHz & This work\\
        \bottomrule
    \end{tabular}
    }
\end{center}
\caption{Typical parameters comparison of Er-doped TFLN lasers}
\label{Table. 1}
\end{table}

The mode splitting and mode competition could be seen throughout the experiment, which is detrimental for high-slope-efficiency and single-mode laser output. The high-power pump laser in the cavity would couple with high-order modes, hindering the continuous increasing of fundamental mode power. The single mode laser output with good monochromaticity, high stability and good beam quality is preferred in photonic integrated circuit. To this end, many methods could be utilized. For example, replacing the microdisk cavity with a microring of a proper width could get a single transverse mode. Further, smaller radius brings larger free spectral range (FSR) to reduce the longitudinal modes at target wavelength band. The Vernier effect in which two coupled cavities with different FSRs could also achieve single mode laser. Besides, the thermal effect of the device shows another method to get single mode emission.
980 nm is another pump wavelength to excite Er ions, while in our experiment the pulley coupler between the bus waveguide and the microdisk cavity is not suitable for efficient coupling of such pump laser. Considering the wavelength division of pump laser and signal laser, 980-nm pump might be a better choice. The net internal gain of a single spiral waveguide amplifier pumped by 980-nm laser is larger than that by 1480-nm laser, which is promising for an integrated device with higher slope efficiency.

In summary, by integrating a microdisk cavity with a spiral waveguide on Er-doped TFLN, we demonstrate an efficient on-chip light source with high on-chip laser power (7.989 µW), high slope efficiency (0.43$\%$) and narrow linewidth (47.86 kHz). The performance of our device is characterized thoroughly and show superiority over the conventional microcavity laser. By applying thermal control, a tunable laser is obtained, and a single mode lasing is observed. The device has a potential to be monolithically integrated with other functional device to form a complete PIC. The improved efficiency also provide solution for quantum applications. By selective doping, the absorption loss brought by Er ions could be avoided, and by efficient on-chip filter, the signal laser could propagate to the following devices for modulating, frequency-doubling, or other application.

\section*{Method}

\subsection*{Device fabrication}

\begin{figure}[h]
\centerline{\includegraphics[width=0.8\textwidth]{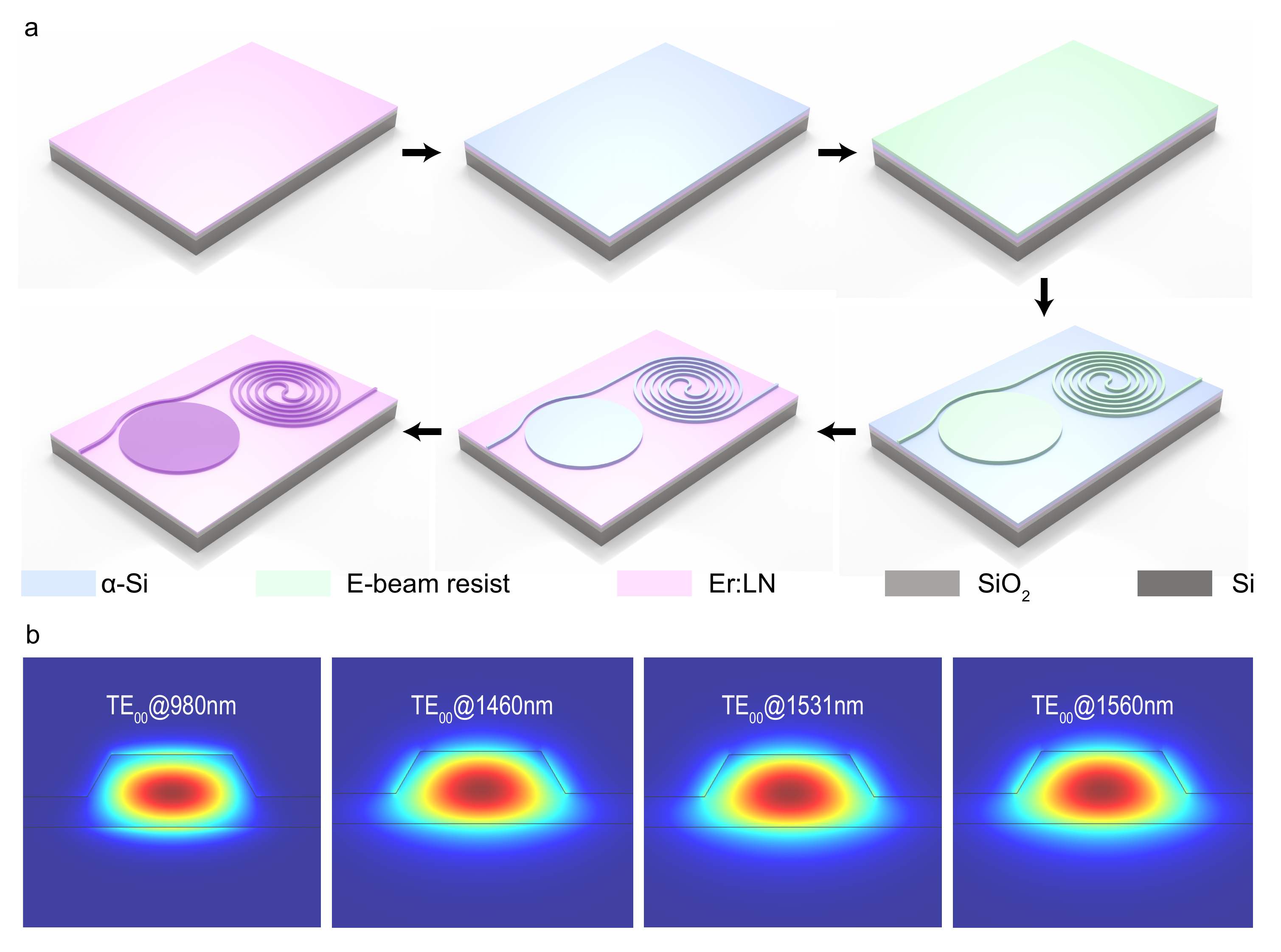}} 
\caption{\textbf{Schematic of the fabrication of the IAL and mode profile in the waveguide.} \textbf{a} Brief fabrication process of the IAL. \textbf{b} Simulated electric field distribution of the TE$_{00}$ modes at 980 nm, 1460 nm, 1531 nm and 1560 nm.} 
\label{Fig. 6} 
\end{figure}

The device is fabricated on a 1-mol$\%$ Z-cut Er-doped TFLN, with 600-nm thick Er-doped LN (Er:LN), 2-µm-thick silica (SiO$_2$) and 500-µm-thick silicon (Si) substrate. Here, we select the LN as a host material due to its excellent physical properties compared to other materials. In order to obtain uniform Er ions doping concentration and achieve a better gain effect, we doped Er ions into LN during the crystal growth processes \cite{liu2021chip}. The brief fabrication processes of the IAL (shown in Fig. \ref{Fig. 6}a) mainly includes six steps: (1) a 600-nm thick amorphous silicon was deposited as a hard etching mask; (2) a layer of resist was spin-coated onto the Er:TFLN; (3) the spiral waveguide structure was patterned via electron- beam lithography (EBL); (4) the mask layer patterns was transferred to the Er:LN layer with an Ar$^+$ plasma etching process; (5) the residual mask was removed by wet etching; the two edge facets were milled by focused ion beam (FIB). Fig.\ref{Fig. 6}b shows the simulated electric field distribution of the optical mode at the mentioned wavelengths.

\subsection*{Experimental setup}

The experimental setup is shown in Fig. S3. A tunable or non-tunable pump laser at around 1480 nm/1460 nm, a C-band tunable laser, a polarization controller, a waveguide-fiber coupling system, several beam splitters/combiners and an optical spectrum analyzer are used to characterize the IAL, microdisk laser and spiral waveguide amplifier. The laser linewidth is obtained through the beat note signal of the IAL and a C-band tunable laser using an electrical spectrum analyzer.

\section*{Acknowledgement}
The authors thank Dr. Jin Li and Prof. Chun-Hua Dong from University of Science and Technology of China for their technical support. The authors thank the Center for Advanced Electronic Materials and Devices (AEMD) of Shanghai Jiao Tong University (SJTU) for the supports in device fabrications. This work was supported by the National Natural Science Foundation of China (Grant Nos. 12134009), and SJTU No. 21X010200828.

\section*{Author details}

$^1$School of Physics and Astronomy,  State Key Laboratory of Advanced Optical Communication Systems and Networks, Shanghai Jiao Tong University, 800 Dongchuan Road, Shanghai 200240, China;  $^2$School of Physics, Ningxia University, Yinchuan, 750021, China; $^3$Shanghai Research Center for Quantum Sciences, Shanghai 201315, China; $^4$Collaborative Innovation Center of Light Manipulations and Applications, Shandong Normal University, Jinan 250358, China.

\section*{Author contributions}

J.W. X.Y. and Y.C. prepared the manuscript in discussion with all authors. X.Y. designed and fabricate the devices. J.W. X.Y. and Y.C. performed the experiments and analyzed the data. X.W. provided help in numerical simulation. Y.C. X.W., T.Y. and C.C. revised the manuscript. H. L. helped with the fabrication process. Y.C. and X.C. supervised the project.

\section*{Conflict of interest}

The authors declare no competing interests.

\bibliography{sample}
\bibliographystyle{unsrt}

\end{document}